\documentclass[12pt,a4paper]{article}
\usepackage[preprint]{dfttrob}
\usepackage{times}
\usepackage[fleqn]{amsmath}
\usepackage{graphicx}
\usepackage{amssymb}
\dfttnum{DFTT 29/2001}

\newcommand{\pt}{\ensuremath{p_{\mathrm{t}}}}

\newcommand{\ee}{\mbox{e$^+$e$^-$}}

\newcommand {\mass} {\mbox{\rm GeV$\kern-0.15em /\kern-0.12em c^2$}}
\newcommand {\tev} {\mbox{${\rm TeV}$}}
\newcommand {\gev} {\mbox{${\rm GeV}$}}

\newcommand {\mom} {\mbox{\rm GeV$\kern-0.15em /\kern-0.12em c$}}

\newcommand {\gmom} {\mbox{\rm GeV$\kern-0.15em /\kern-0.12em c$}}
\newcommand {\mmass} {\mbox{\rm MeV$\kern-0.15em /\kern-0.12em c^2$}}
\newcommand {\mmom} {\mbox{\rm MeV$\kern-0.15em /\kern-0.12em c$}}

\newcommand{\Jpsi} {\mbox{J\kern-0.05em /\kern-0.05em$\psi$}\xspace}

\newcommand{\ppbar}{\mbox{$\mathrm {p\overline{p}}$}}

\newlength{\digitwidth} 
\newlength{\ql} \newlength{\qll} \newlength{\qlll}  \newlength{\qllll}
\newlength{\qlp} \newlength{\qllp} \newlength{\qlllp} \newlength{\qllllp}
\newlength{\qlup} \newlength{\qlupp}
\begin{document}

\providecommand{\Alice}{ALICE}
\providecommand{\SpS}{Sp\=pS}
\providecommand{\nbar}{{\bar n}}
\providecommand{\ksoft}{k_{\text{soft}}}
\providecommand{\nsoft}{\nbar_{\text{soft}}}
\providecommand{\ksemi}{k_{\text{semi-hard}}}
\providecommand{\nsemi}{\nbar_{\text{semi-hard}}}
\providecommand{\asoft}{\alpha_{\text{soft}}}
\providecommand{\avg}[1]{\langle #1 \rangle}

\title{Global event properties in proton-proton physics with ALICE}
\author{A. Giovannini and R. Ugoccioni\\
Dipartimento di Fisica Teorica and I.N.F.N. -- sezione di Torino\\
via P. Giuria 1, 10125 Torino, Italy}
\maketitle

\begin{abstract}
\Alice\ is a unique opportunity for studying low \pt\ physics and
minimum bias events, and consequently for hunting substructures
in strong interactions.
The general question concerning global event properties in pp 
and \ppbar\ physics
is indeed whether substructures can be seen in the
data in a model-independent way.
In other terms, can we identify the interface between perturbative and
non-perturbative regimes?
\end{abstract}
\vfill
\begin{center}
Contribution to the ALICE Physics Performance Report, to appear.
\end{center}
\vspace*{2cm}

\newpage
\section{Introduction}
\Alice\ is a unique opportunity for studying low \pt\ physics and
minimum bias events, and consequently for hunting substructures
in strong interactions.
The general question concerning global event properties in pp 
and \ppbar\ physics
is indeed whether substructures can be seen in the
data in a model-independent way.
In other terms, can we identify the interface between perturbative and
non-perturbative regimes?

A successful approach to this question has recently emerged
\cite{TH:Torino2000}, as will be shown in the following.

\section{A summary of substructure hunt in pp collisions}
\label{TH:A_summary_of_substructure_hunt}

What is known on the subject comes from data taken at ISR, at \SpS\
collider (UA1, UA2 and UA5 experiments) and the  Tevatron collider
(CDF and E735 experiments) as well as from dedicated theoretical work.

\subsection{Energy widening of multiplicity distributions.}
\label{TH:Energy_widening_of_MD}
\begin{figure}
  \begin{center}
  \mbox{\includegraphics[width=\textwidth]{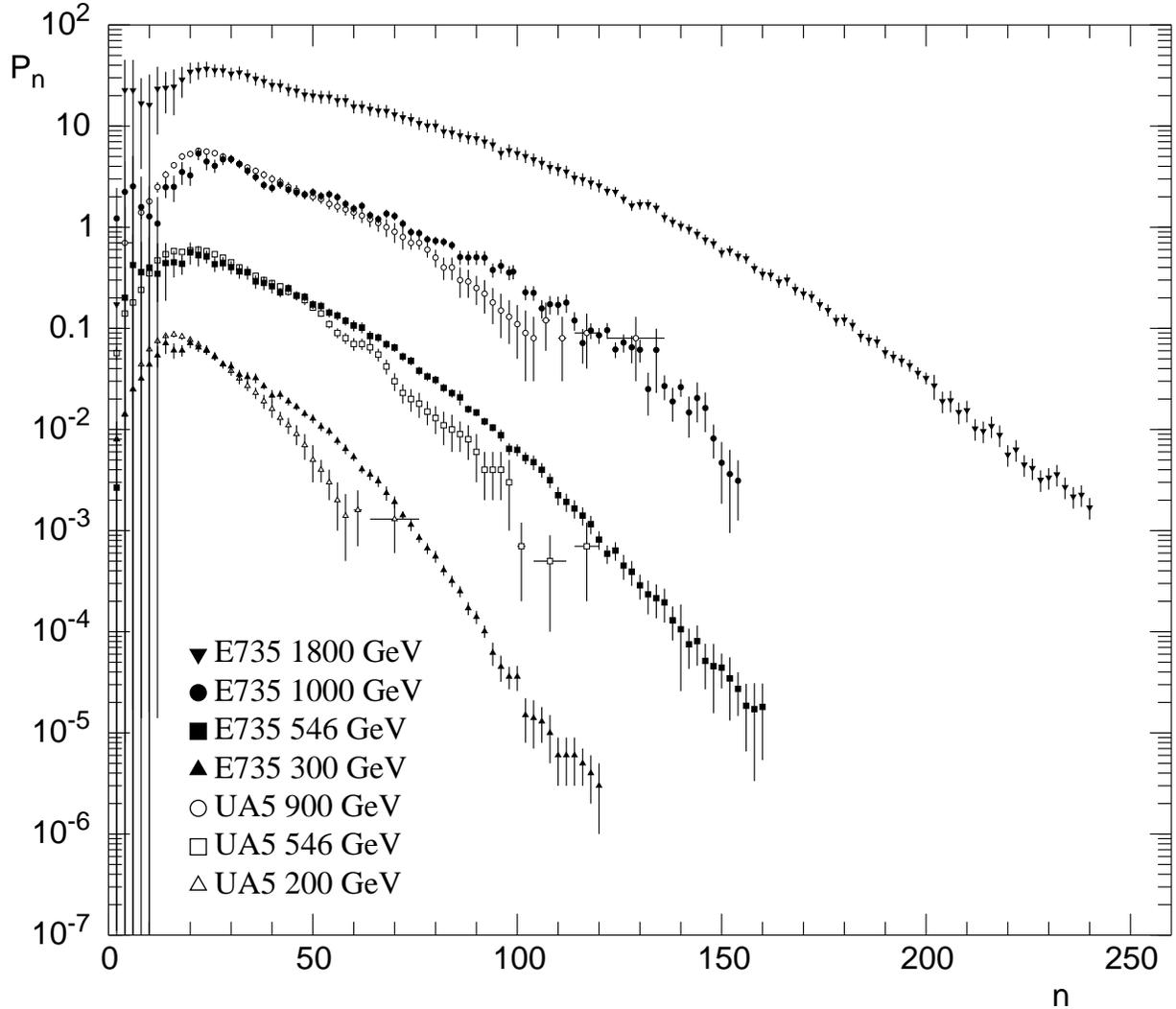}}
  \end{center}
  \caption{E735 results on charged particle multiplicity distributions
  in full phase space compared with UA5 results
  \cite{TH:walker}. 
  Data from the two experiments which were taken at nearly the same
  energy are rescaled by the same factor.
  }\label{TH:fig:MD_UA5E735} 
  \end{figure}

It is to be stressed that recent results from Tevatron
(E735 Collaboration, \cite{TH:walker}) 
on full phase space multiplicity distributions
do not completely agree with those obtained at comparable energies
at the \SpS\ collider (UA5 Collaboration, 
\cite{TH:UA5:rep,TH:UA5:3}),
see Fig.~\ref{TH:fig:MD_UA5E735}. Tevatron data are more precise than
\SpS\ data at larger multiplicities (they have larger
statistics and extend to larger multiplicities than UA5 data), but
much less precise at low multiplicity.
Both sets of data show a shoulder structure, but the Tevatron MD is
somewhat wider. It should be noticed that E735 data are measured
only in $|\eta|<3.25$ and $\pt > 0.2$~\gmom\ 
then extended to full phase space via a Monte
Carlo program.

A standing problem!

\subsection{The fit with two negative binomial distributions.}
\label{TH:The_fit_with_two_NBD}
The negative binomial (NB) behaviour  
for final charged particles multiplicity
distributions can be trusted  in hadron-hadron collisions in full phase 
space only up to ISR energies \cite{TH:giacomelli}. 
At higher  energies shoulder  structures  
start to  be clearly visible as shown by the UA5 Collaboration at CERN
\ppbar\ collider \cite{TH:UA5:3}.
The idea firstly suggested by C. Fuglesang  \cite{TH:Fug} is  
to explain observed 
NB regularity violations as  the effect of the  weighted superposition
of two classes of events, the 
multiplicity distribution of each component being of NB type:
\begin{equation}
P_n(\asoft; \nsoft, \ksoft; \nsemi, \ksemi) =
     \asoft P_n^{\text{NB}}(\nsoft, \ksoft) +
    (1 -\asoft) P_n^{\text{NB}}(\nsemi, \ksemi)  ;   
	\label{TH:eq:combo}
\end{equation}
where, for each class, 
$\nbar$ is the average multiplicity and parameter $k$ is linked
to the dispersion $D$ by $k^{-1} = D^2/\avg{n}^2 - 1/\avg{n}$.
The two classes are interpreted as soft events (events without
mini-jets)  and semi-hard events (events with mini-jets) and
consequently the weight $\asoft$ is  the fraction of events without mini-jets
as measured by UA1 \cite{TH:UA1:minijets}.
The conclusion is that the 
proposed fit in terms of  the superposition of two negative
binomial multiplicity distributions (NBMD's) is quite good.

\subsection{Extrapolations to LHC energy.}
\label{TH:Extrapolations_to_LHC_energy}
The point is to find  acceptable   energy dependence of the NB parameters  
$k$ and $\nbar$ for the two components substructures and the corresponding
weight factor $\asoft$. In a region where QCD has no predictions one must 
proceed by phenomenological assumptions 
\cite{TH:combo:prd,TH:combo:eta}, 
which are summarised in the following.

The first assumption concerns energy dependence of the total average charged 
particle multiplicity \cite{TH:combo:prd}:
\begin{equation}
	\nbar = 3.01 - 0.474 \ln\sqrt{s} + 0.745 \ln^2\sqrt{s} .  
		\label{TH:eq:nbar}
\end{equation}
The $\ln^2s$ term is interpreted as the effect of the sharp increase
of minijet production and/or double (even triple) parton collisions as
suggested by the rapid increase of $\avg{\pt}$ with multiplicity.
Since below 200~\gev\ c.m.\ energy one single NB describes multiplicity
distribution  
data very well and above 200~\gev\ c.m.\ energy the soft component has been
disentangled, it is proposed to extrapolate the logarithmic 
increase with energy of the average charged particle multiplicity  of the
soft component,  $\nsoft$, also at higher c.m.\ energy.
From $\nbar$ and $\nsoft$, one can calculate the behaviour of
$\asoft$, which, using the UA1 result $\nsemi \approx 2 \nsoft$, gives
$\asoft \approx 0$ at 100~\tev, but, if one allows again a $\ln^2 s$
term in $\nsemi$
(with a small coefficient, of the order of 0.1, it is also consistent
with the data) one still gets at that energy a sizable (20\%)
contribution from soft events.

The second assumption concerns the energy dependence of the
NB parameter $k$. For the soft component, KNO scaling was assumed, 
since it was found to hold in the ISR region. For the semi-hard
components three scenarios were examined, two extreme ones (scenario 1 in
which the semi-hard component also obeys KNO scaling and scenario 2 in
which the semi-hard component violates KNO scaling with $1/k$ growing
linearly with $\ln s$) and an intermediate case (scenario 3,
where $1/k$ grows not as fast
as linearly with a shape inspired by pQCD calculation). 

\begin{figure}
  \begin{center}
  \mbox{\includegraphics[width=0.7\textwidth]{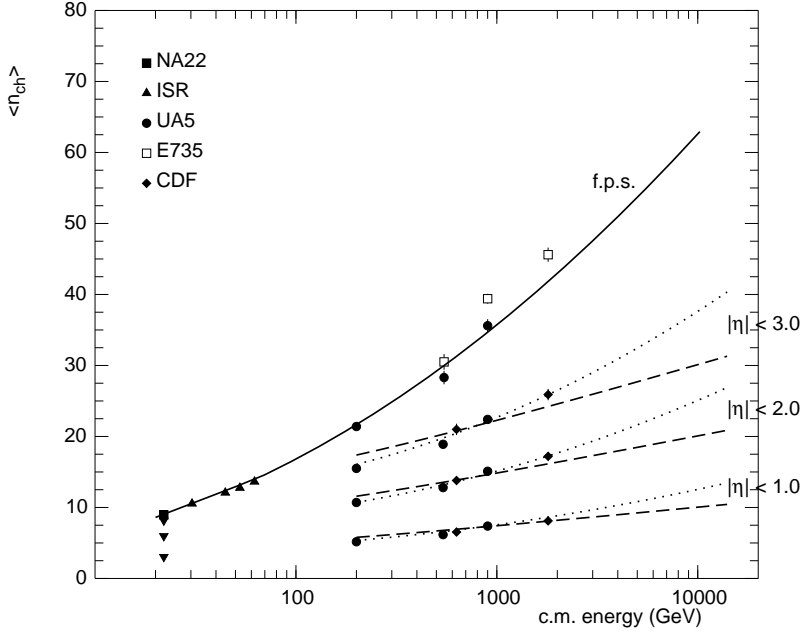}}
  \end{center}
  \caption{Average charged multiplicity in full phase space
	and in three pseudo-rapidity interval vs.\ c.m.\ energy. 
	Experimental data shown refer to non-single-diffractive data. The solid
  line shows Eq.~(\ref{TH:eq:nbar}), the dashed and dotted lines
	are described in the text.}\label{TH:fig:ncharged}
  \end{figure}

\begin{figure}
  \begin{center}
	\mbox{\includegraphics[width=0.8\textwidth]{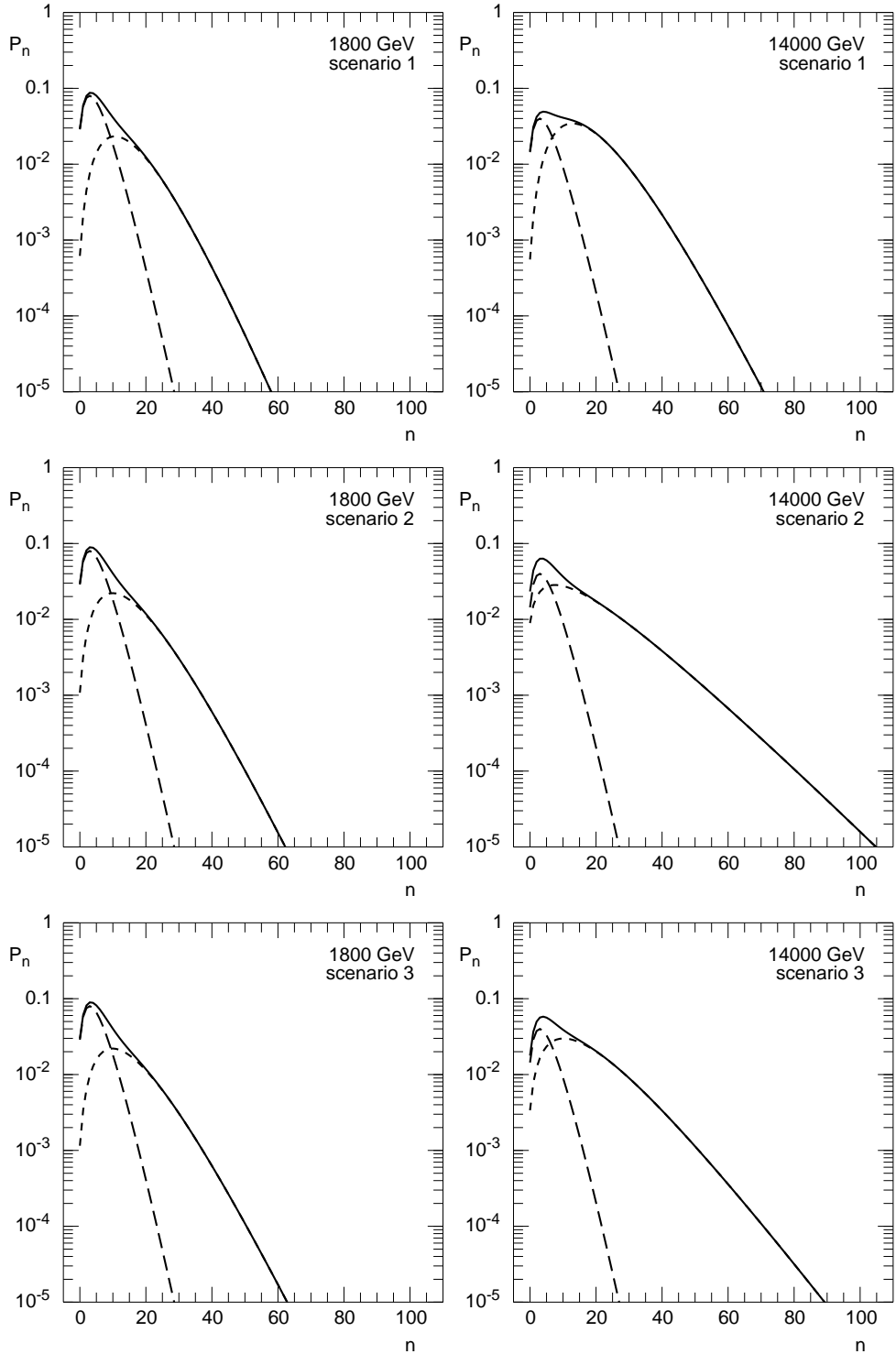}}
  \end{center}
  \caption{Predicted charged particle multiplicity distributions in
  the two component model \cite{TH:combo:eta} for 
  the interval $|\eta|<1$
  at Tevatron and LHC energies, where the two components (dashed
  lines) are scaled by $\asoft$ and $1-\asoft$ respectively,
  Eq.~(\ref{TH:eq:combo}), with $\asoft=0.6$ at 1800~\gev\
	and 0.3 at 14~\tev.}\label{TH:fig:md_GU}
  \end{figure}

The behaviour of  $\nbar$   in the TeV region  based on UA5 data
(Eq.~(\ref{TH:eq:nbar})) is shown in 
Fig.~\ref{TH:fig:ncharged} (solid line).  
It is compared with the mentioned E735 results \cite{TH:walker}
obtained by extrapolating to full phase space experimental data
with a Monte Carlo calculation; observed discrepancies 
are clearly consequences of the differences noticed  between E735  
results and UA5 data on charged particle multiplicity distributions shown
in Fig.~\ref{TH:fig:MD_UA5E735} and discussed in 
Section~\ref{TH:Energy_widening_of_MD}.
Accordingly, one relies on Eq.~(\ref{TH:eq:nbar}) predictions.

Coming to particle rapidity density it should be pointed out that by assuming 
only a longitudinal growth of  phase space and constant height of the 
rapidity  plateau with c.m.\ energy for semi-hard events, as done in 
Ref.~\cite{TH:combo:eta}, 
CDF data \cite{TH:CDF:dNdeta} in pseudorapidity 
intervals  are underestimated (see Fig.~\ref{TH:fig:ncharged}, 
dashed lines). 
These data are well
described by allowing a $\ln^2 s$ growth of the total rapidity plateau
(Fig.~\ref{TH:fig:ncharged}, dotted lines):
from this consideration one deduces a  more appropriate growth of
the semi-hard plateau height; the constraint is that $\nsemi$
in full phase space follows a logarithmic growth with $\sqrt{s}$ as
discussed above.
Predicted charged particle multiplicity  distributions 
in the three scenarios of the two component model \cite{TH:combo:eta}
for the interval  $|\eta| < 1$ at Tevatron and LHC energies, calculated
for the last mentioned case, are shown 
in Fig.~\ref{TH:fig:md_GU} (notice
that the comparison with CDF data, in view of the relatively large
value of their resolution $\pt > 0.4$~\gmom, is questionable).
If this behaviour for the semi-hard component will be confirmed  by data
one should conclude  that semi-hard events populate mainly the central 
rapidity region giving an important contribution to the increase of
charged particle density in central rapidity intervals.

\begin{figure}
 \begin{center}
 \mbox{\includegraphics[width=0.4\textwidth,height=6.5cm]{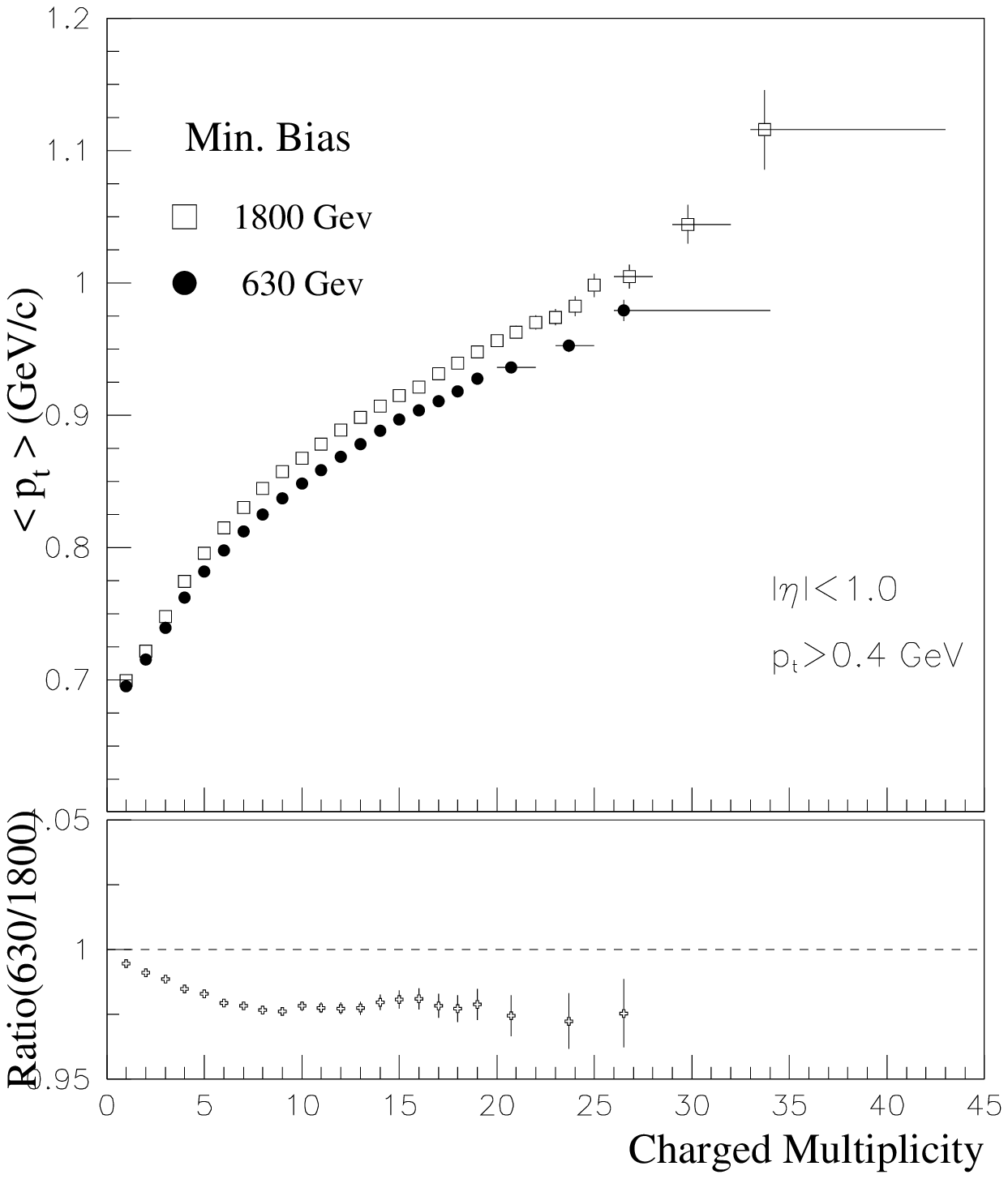}}\\
 \mbox{\includegraphics[width=0.4\textwidth,height=6.5cm]{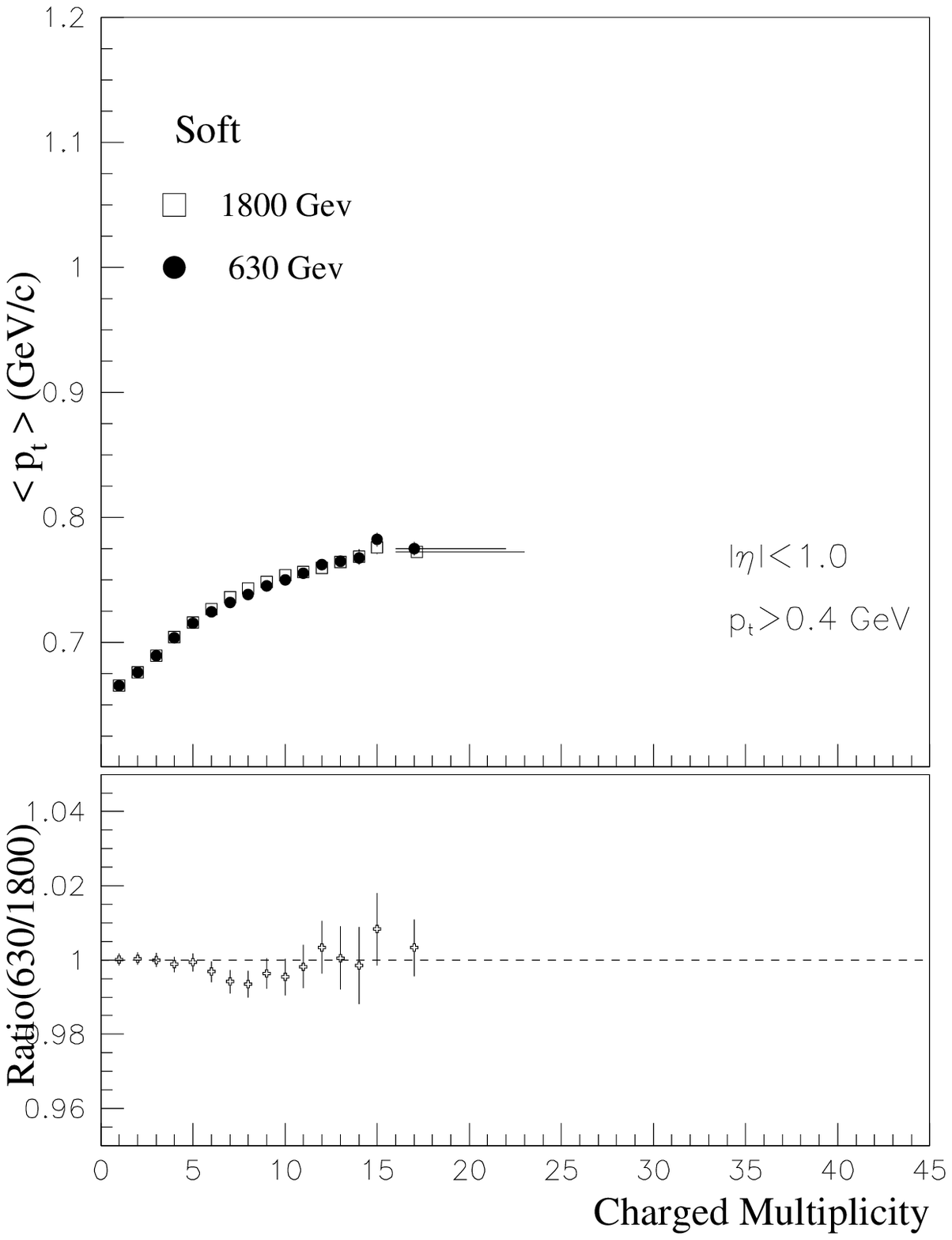}}~
 \mbox{\includegraphics[width=0.4\textwidth,height=6.5cm]{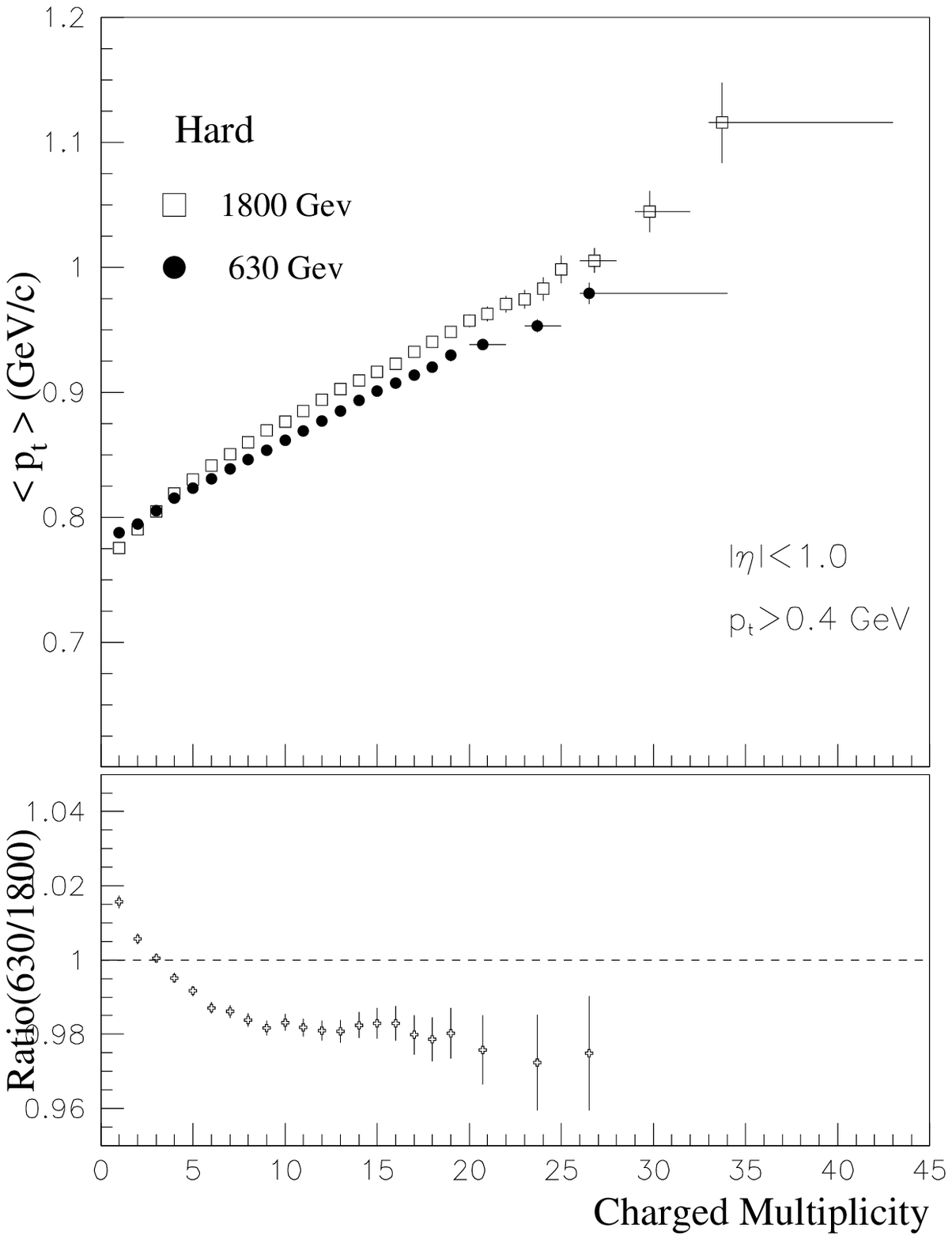}}
 \end{center}
	\caption{CDF measurements on average transverse momentum vs
 multiplicity in the full sample and in different classes 
 of events (with and without minijets, called
 resp. `hard' and `soft' in this figure) \cite{TH:CDF:soft-hard}}
		\label{TH:fig:cdf_pt_mult}
 \end{figure}

\subsection{Soft and hard samples at Tevatron.}
\label{PPGLOABL:Soft_and_hard_samples_at_Tevatron}
It was found by CDF \cite{TH:Rimondi:Torino2000}
that by subdividing the minimum bias
sample into two groups, characterised respectively by the absence
(`soft' events) or the presence (`hard' events) of mini-jets,
interesting features of the reaction can be investigated.
More precisely, a `hard' event has been defined as an event with at least
one calorimeter cluster in $|\eta|<2.4$, a cluster being defined as a
seed calorimeter tower with at least 1~\gev\ transverse energy $E_t$ plus at
least one contiguous tower with $E_t \ge 0.1$~\gev.  
A subdivision which is interesting per se and can be tested at 14~\tev.

In summary, the soft component is found to satisfy KNO scaling
(as expected in \cite{TH:combo:eta}), while
the hard one does not; also the $\avg{\pt}$ distribution scales at
fixed multiplicity in the soft component and not in the hard one;
the dispersion of $\avg{\pt}$ vs.\ the inverse of the multiplicity is
compatible with an extrapolation to 0 as $n\to\infty$ in the soft
component but not in the hard one, indicating 
in this limit a lack of correlations in
the soft component.

The correlation between $\avg{\pt}$ and
multiplicity was explained by UA1 \cite{TH:UA1:pt_nch}
as related to the onset of gluon radiation.
It should be noticed that at CDF  such a correlation is found 
to some extent for both the soft and the
hard subsamples  as shown in Fig.~\ref{TH:fig:cdf_pt_mult},
but the soft subsample is seen to start to saturate
(CDF data shown have $|\eta|<1.0$ and $\pt > 0.4$~\gmom).

\subsection{Low and high transverse momentum at \SpS.}
\label{TH:Low_and_high_transverse_momentum}
An investigation has been recently carried out
\cite{TH:Buschbeck:Torino2000} on UA1 data also
based on the superposition idea:
particles are selected according to their \pt\
(this selection should be contrasted with that performed by CDF in
terms of classes of events).
It is seen that the high-\pt\ sample ($\pt > 0.7$~\gmom)
behaves very differently from the low-\pt\ sample.
This difference is interpreted as the effect of a more intense
jet-like activity in the high-\pt\ sample.
As far as the low-\pt\  sample is concerned, it was shown 
that the
dependence of the correlation strength and of higher order cumulants
on multiplicity is important in order to test different theoretical
models (Monte Carlos are totally inadequate here): for this task the
low-\pt\ cut-off of \Alice\ at LHC is required.
In addition, it would be very interesting to analyse in this way the
soft and semi-hard components.

\begin{figure}
  \begin{center}
  \mbox{\includegraphics[scale=0.55]{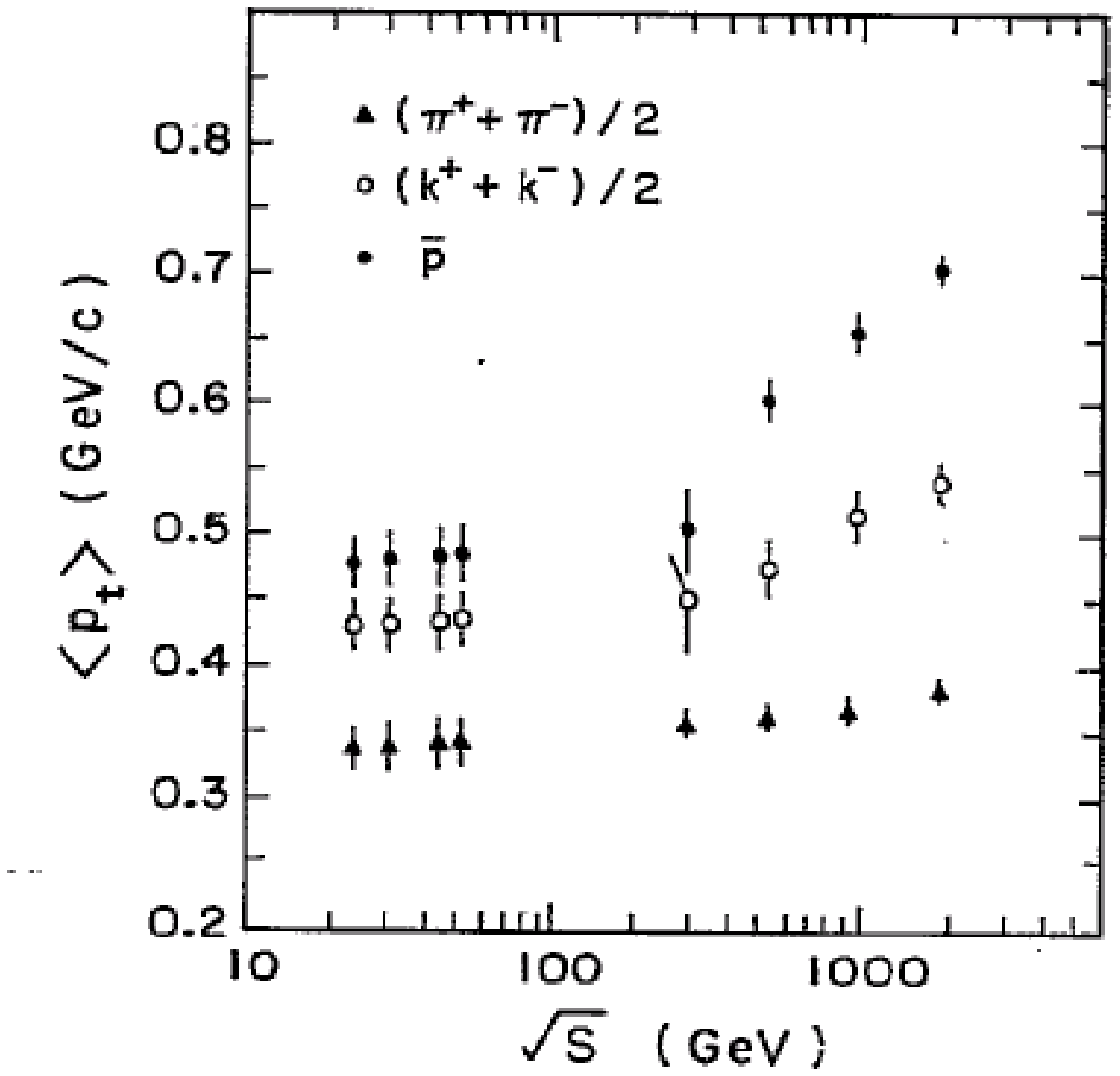}}
  \mbox{\includegraphics[scale=0.45]{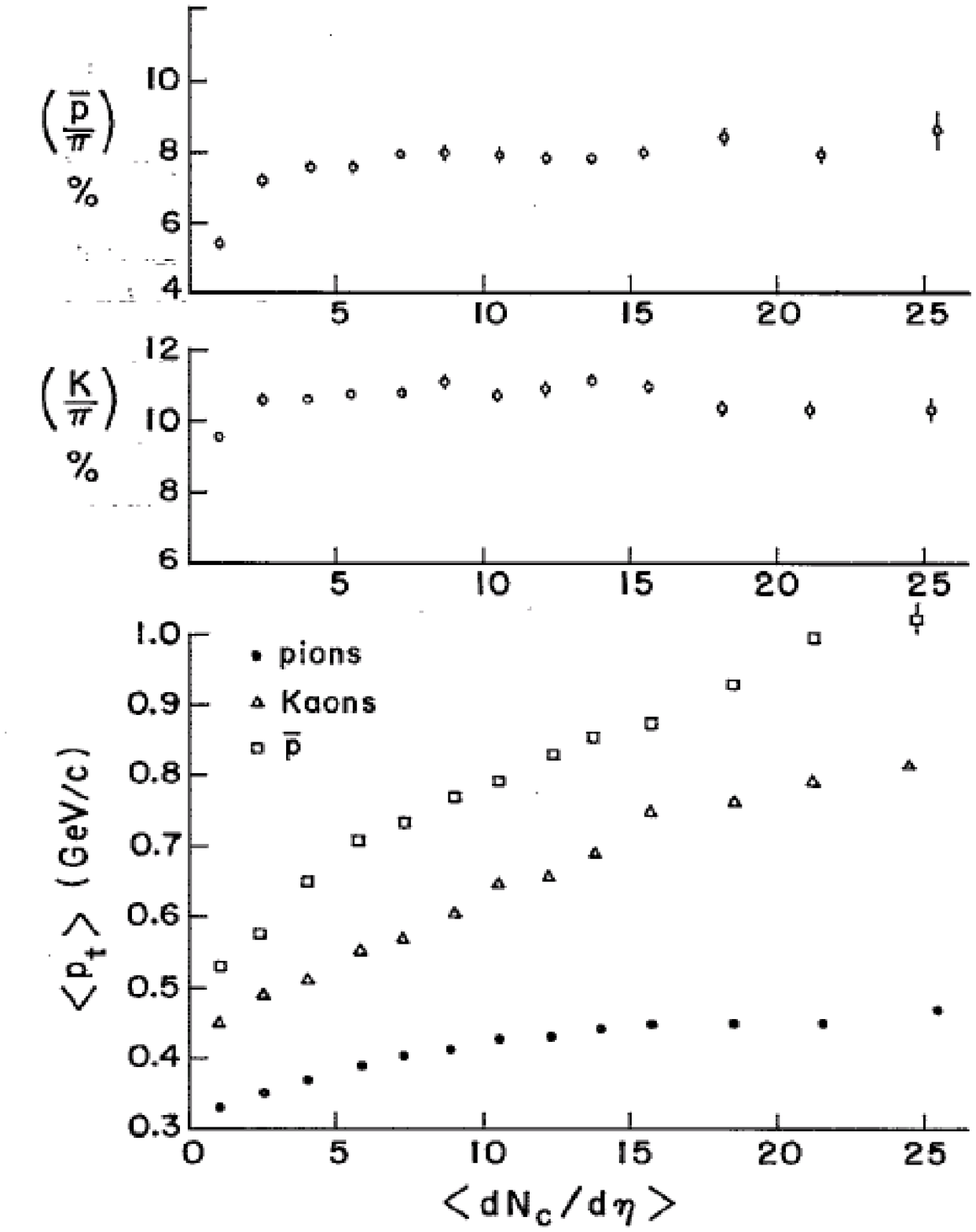}}
  \end{center}
  \caption{Measurements by E735 \cite{TH:Alexopoulos:1993wt} of average
  \pt\ for identified particles vs.\ c.m.\ energy and vs.\ particle
  density. The relative abundance of particles is also shown as a
  function of particle density.}\label{TH:fig:e735_pt}
  \end{figure}

\subsection{Average transverse momentum versus multiplicity.}
\label{TH:Average_transverse_momentum_versus_multiplicity}
The study should be mentioned at Tevatron (E735) 
\cite{TH:Alexopoulos:1993wt} of the
correlation between $\avg{\pt}$ and multiplicity done separately for
pion, kaons and antiprotons, which shows that the behaviour is rather
different, as illustrated in
Fig.~\ref{TH:fig:e735_pt}, and still not theoretically understood. 
Data show saturation in \pt\ at large
multiplicity for $\pi^{\pm}$, not for and $K^{\pm}$ and $\bar p$
(but for kaons it saturates if a cut is imposed on momenta
$\pt < 1.5$~\gmom, see \cite{TH:Alexopoulos:1990pt}).
At LHC, with $10^9$ events, the predictions discussed
in Section~\ref{TH:Extrapolations_to_LHC_energy} allow the possibility
to reach densities from 40 (scenario 1) to 60 (scenario 2) particles
per unit rapidity, with good statistics (1000 events).
This fact has far reaching consequences which can be tested with \Alice\
especially at low momentum and the study can be extended to baryon
production in the central region (Alice can also measure $\Lambda$).
All this, plus the relative abundance of particle species, carries precious
information on possible quark-gluon plasma formation, in particular in view of
such properties as strangeness enhancement and baryon stopping
(see elsewhere in this Chapter).

\section{Investigations in pp collisions with ALICE}
\label{TH:Investigations_in_pp_collisions}

\subsection{Shape of the multiplicity distribution.}
The average multiplicity
$\avg{n}$ grows with the c.m.\ energy $\sqrt{s}$: the best fit
involves a term proportional to  $\ln^2 s$, 
whose theoretical basis is not yet understood.
Most theoretical works predict in fact a power-law in $s$ or a linear rise
with $\ln s$.  Is LHC energy
large enough to distinguish these behaviours? Purely statistical
extrapolation based on the above mentioned scenarios 
(Section~\ref{TH:Extrapolations_to_LHC_energy}) show that the average
multiplicity in full phase space can be measured with 0.2\% error
with $10^5$ events.

The ratio $D/\avg{n}$, where $D$ is the dispersion,
is constant if KNO scaling holds. 
KNO scaling is an asymptotic prediction:
data at ISR energies are compatible with an `early' KNO scaling, but
\SpS\ data clearly showed a violation.
Purely statistical extrapolation based on 
the above mentioned scenarios shows that the variance $D^2$
can be measured with less than 1\% error
with $10^5$ events both in full phase space and 
within the \Alice\ acceptance. 

\subsection{Shape fits.}
It will be possible to verify the extent of forward-backward (FB)
correlations
seen at \SpS\ and at Tevatron, to distinguish whether FB
correlations grow or decrease and to study their link with the MD
\cite{TH:Ekspong}.

Predictions exist for the multiplicity distribution
at 14~\tev, (e.g.\ \cite{TH:combo:prd,TH:combo:eta}
and \cite{TH:Kaidalov:PPR}; an alternative point of view
on the two-component structure based on impact parameter analysis
is presented in \cite{TH:two-comp}). 
As previously explained in 
Section~\ref{TH:Extrapolations_to_LHC_energy}, 
in \cite{TH:combo:prd,TH:combo:eta}
three scenarios can be distinguished at LHC
by the value of $k^{-1} \approx D^2/\avg{n}^2$. 
It turns out that both in full phase space and in
restricted pseudo-rapidity regions 
the difference between the scenarios can be 
sharply defined if $D^2/\avg{n}^2$ is measured to 15\% accuracy,
as seen from the following table:
\begin{center}
  \begin{tabular}{|r|c|c|c|}	\hline
                 & \multicolumn{3}{c|}{$k^{-1}$} \\
	scenario & f.p.s. & $|\eta|<1.5$ & $|\eta|<1$ \\ \hline
  \hline
	       1 & 0.17   &  0.37      & 0.39 \\ \hline
	       2 & 0.42   &  0.74      & 0.78 \\ \hline
	       3 & 0.24   &  0.54      & 0.56 \\ \hline
  \end{tabular}
\end{center}
Purely statistical extrapolation shows that in these scenarios 
$10^5$ events will yield 0.5\% error, which is expected to be good
enough to give extremely relevant information even if none
of the envisaged scenarios turns out to be adequate (this would be
a strong indication of the presence of an anomalous additional (hard?)
component).

The scenarios and predictions should be confirmed not only by the
analysis of the first two moments: the whole distribution should be
involved. Purely statistical extrapolation shows that data produced in
one scenario cannot be adequately fitted by either one of the other
scenarios already with $10^5$ events. Also fits based on one component 
only (of NB type) can be excluded with such statistics, as well as
the other predictions.

Again on the global side, the information entropy $S = -\sum_n
P_n \log P_n$ can be calculated and compared with 
the behaviour assumed in \cite{TH:SimakSumbera}
and the one implicit in \cite{TH:combo:prd,TH:combo:eta}.
Purely statistical extrapolations based on the latter reference show that 
information entropy can be measured to better than 0.1\%,
both in full phase space and in $|\eta|<1$, with $10^5$ events, 
which is more than enough to distinguish again the mentioned
scenarios among them and from the expectation 
\cite{TH:SimakSumbera} of a scaling
with available phase space. Incidentally, the latter option would 
produce a correction to $\avg{n}$ of a few percent,
thus perfectly identifiable by LHC already with $10^5$ events
(ignoring again systematic errors).

Feynman scaling ($dN/dy$ around $y = 0$ independent of $\sqrt{s}$,
where $y$ is the rapidity), introduced initially at ISR energies,
has been shown to be violated
at higher energies: as discussed previously in 
Section~\ref{TH:Extrapolations_to_LHC_energy},
the violation is due not only to the increasing importance of semi-hard 
events, but one expects a growth also of the semi-hard plateau; 
this should be checked at LHC. 

The $E d^3N/dp^3$ vs.\ \pt\
distribution is interesting because it can be compared with
results from Tevatron and \SpS.
Is it possible to identify two (or more)
components (e.g., different slopes) in this
distribution, even if the soft events have
a much smaller \pt\ than the hard ones? The CDF collaboration
has shown that the single event's average \pt\ is a good candidate
for event classification, especially in combination with the event
multiplicity (see again Fig.~\ref{TH:fig:cdf_pt_mult}).
Multiplicity classes can be defined
and whole distributions (e.g.\ in transverse momentum) can be looked at 
as a function of multiplicity or particle density
 \cite{TH:Rimondi:Torino2000}.
Very interesting information for investigating the mechanism of
particle production can come from the study of identified particles.

\subsection{Sign oscillations of higher order moments.}
$H_q$ moments (the ratio of factorial, $F_q$, to 
cumulant, $K_q$, moments) 
are very important in defining substructures in the multiplicity
distribution \cite{TH:hqlett:2}, as experimentally
measured by L3 at LEP \cite{TH:L3:mangeol}, 
but require very large statistics in order to be measured. 
Purely statistical extrapolations 
show that, in $|\eta|<1$, $10^5$ events
allow to confidently calculate $H_q$ up to order $q = 7$, 
but the expected $10^9$ events,
will allow to reach at least order $q = 14$, quite adequate
for the purpose.
In fact $H_q$ vs.\ $q$ oscillations in hadron-hadron collisions in the \gev\
region \cite{TH:hqlett:2,TH:combo:prd}  
and in \ee\ annihilation at LEP energies 
\cite{TH:L3:mangeol} have been explained 
as the effect  of the weighted  superposition of classes of different
topology, each class being described by a NBMD with different parameters.
Observed shoulder structure in final charged particle multiplicity 
distributions and $H_q$ vs.\ $q$ oscillations  have  in this framework the same
origin. In addition being  $H_q$ the  ratio of factorial  to cumulant moments,
$F_q/K_q$, the occurrence of the NBMD for a sound description of $H_q$
behaviour in each substructure via generalised local parton-hadron duality
(GLPHD), i.e.\ the statement that $F_q$ at hadron level is equal to
$\rho^q  F_q$ at parton level,  with $\rho$ defined by  
$\nbar_{\text{hadron}} = \rho \nbar_{\text{parton}}$,
leads to the conclusion that 
$H_q$ at hadron level coincide with $H_q$ at parton level: an interesting
possibility to understand deeply QCD and/or GLPHD.

\subsection{Underlying event for Higgs production.}
As Bjorken pointed out \cite{TH:Bjorken2}, at LHC energy
processes will happen characterised 
by the presence of virtual electroweak bosons in the hard subprocess,
like WW scattering via W or even Higgs exchange, 
with the bosons treated as partons of the proton beam.
If the W's then decay leptonically, a feature of the
event will be a large rapidity gap, i.e., a region without hadrons
separating the beam-jets containing the fragments of the projectiles.
To turn this into a reliable signature, several issues were
raised and discussed in \cite{TH:Bjorken2}: one of them 
demands to know how big the
rapidity gap must be in order that multiplicity fluctuations do not
mimic the effect of $W$ scattering and decay. 
To this question a detailed answer can be given
by studying low multiplicity events: the scenarios previously
described foresee 25 to 35\% of events with less than 4 charged
particles in the central region $|\eta|<1.5$, which seems
a considerable background to the W and Higgs events

One should also not forget that the `void probability' $P_0$, i.e.,
the probability of producing zero charged particles in a given
rapidity interval, has interesting
properties on its own, since in principle its $\nbar$ dependence determines
the full MD \cite{TH:pzero}.

\subsection{Quark-gluon plasma.}
\label{TH:Quark_gluon_plasma}
It has been argued that the increase of $\avg{\pt}$ from ISR to \SpS\
energies and its subsequent flattening with multiplicity could be an
indication of a phase transition; this view has been abandoned and
data analysed later in terms of phase space constraints and the
emergence of mini-jets.
CDF analysis (see Fig.~\ref{TH:fig:cdf_pt_mult}) 
of $\avg{\pt}$ vs.\  central rapidity particle density, 
$dN/dy$, at 630~\gev\ and 1800~\gev\ reveals a 
different behaviour for soft and semi-hard events: a saturation is 
seen in soft events  at $\avg{\pt} \approx 0.5$~\gmom, 
a fact to be contrasted with
the increase from  $\approx 0.44$  up to $\approx 0.7$~\gmom\ 
for (semi)hard events.
In view of the high rapidity density expected  with Alice,  two important 
questions should be asked: 
\textit{a}) does saturation effect for soft events  
continue up to 14~\tev, or,
at a given particle rapidity density,  $\avg{\pt}$ starts to increase
again? In the latter case, could the sharp increase be indicative  of a 
deconfinement transition in hot hadronic matter 
\cite{TH:VanHove:ptQGP}? 
\textit{b}) does  $\avg{\pt}$ for semi-hard events 
continue to increase up to 14~\tev, or 
at a given particle rapidity density  a saturation effect appears? 
One should expect that hard events at some particle
rapidity density  start to appear on top of the semi-hard ones in the 
first case, and the possible occurrence of a phase transition 
in the second one.


\section*{References}
\begin{thebibliography}{110}
\parskip=0.pt \parsep=0.pt \itemsep=0.pt


\bibitem{TH:Torino2000}
{\em Torino 2000: New Frontiers in Soft Physics and Correlations on the
  Threshold of the Third Millennium}, edited by {A. Giovannini and R.
  Ugoccioni} (Nucl.\ Phys.\ (Proc.\ Supp\. ) {\bf B92} February, 2001).

\bibitem{TH:walker}
{T. Alexopoulos et al. (E735 Collaboration)}, Phys.\ Lett.\ {\bf B435}  
  (1998) 453.

\bibitem{TH:UA5:rep}
{G.J.\ Alner et al. (UA5 Collaboration)}, Physics Reports {\bf 154}  
  (1987) 247.

\bibitem{TH:UA5:3}
{R.E.\ Ansorge et al., (UA5 Collaboration)}, Z. Phys.\ {\bf C43}  (1989)  357.

\bibitem{TH:giacomelli}
{G.~Giacomelli and M.~Jacob}, Physics Reports {\bf 55} (1979) 1  .

\bibitem{TH:Fug}
{C.\ Fuglesang},  in {\em Multiparticle Dynamics: Festschrift for L\'eon Van
  Hove}, edited by {A.\ Giovannini and W.\ Kittel} (World Scientific,
  Singapore, 1990), p.\ 193.

\bibitem{TH:UA1:minijets}
{C.~Albajar et al. (UA1 Collaboration)}, Nucl.\ Phys.\  {\bf B309}
  (1988)  405.

\bibitem{TH:combo:prd}
{A.~Giovannini and R.~Ugoccioni}, Phys.\ Rev.\  {\bf D59} (1999) 094020.

\bibitem{TH:combo:eta}
{A.~Giovannini and R.~Ugoccioni}, Phys.\ Rev.\  {\bf D60}   (1999) 074027.

\bibitem{TH:CDF:dNdeta}
{F. Abe et al. (CDF Collaboration)}, Phys.\ Rev.\  {\bf D41}   (1990) 2330.

\bibitem{TH:CDF:soft-hard}
{D. Acosta et al., (CDF Collaboration)}, preprint FERMILAB-PUB-01/345-E,
  FERMILAB.

\bibitem{TH:Rimondi:Torino2000}
{F. Rimondi}, Nucl.\ Phys.\ (Proc.\ Suppl.)  {\bf B92}   (2001) 114.

\bibitem{TH:UA1:pt_nch}
{G. Bocquet et al. (UA1 Collaboration)}, Phys.\ Lett.\ {\bf B366}  (1996)  434.

\bibitem{TH:Buschbeck:Torino2000}
{B. Buschbeck and H.C. Eggers}, Nucl.\ Phys.\ (Proc.\ Suppl.) {\bf B92}
  (2001)  235;
{B. Buschbeck, H.C. Eggers and P. Lipa}, Phys.\ Lett.\ {\bf B481}
  (2000) 187.

\bibitem{TH:Alexopoulos:1993wt}
{Alexopoulos, T. et al. (E735 Collaboration)}, Phys.\ Rev.\  {\bf D48}   (1993) 984.

\bibitem{TH:Alexopoulos:1990pt}
{T. Alexopoulos et al., (E735 Collaboration)}, Phys.\ Rev,\ Lett.\ 
{\bf 64}   (1990) 991.

\bibitem{TH:Ekspong}
{G.\ Ekspong},  in {\em XVI International Symposium on Multiparticle Dynamics},
  edited by {J. Grunhaus} (Editions Fronti\`eres and World Scientific,
  Gif-sur-Yvette and Singapore, 1986), p.\ 309.

\bibitem{TH:Kaidalov:PPR}
{A. Kaidalov}, talk at the Alice PPR meeting, CERN, April 2001.

\bibitem{TH:two-comp}
{J. Dias de Deus and R. Ugoccioni}, Phys.\ Lett.\  {\bf B469}   (1999) 243.

\bibitem{TH:SimakSumbera}
{V. \u{S}im\'ak, M. \u{S}umbera and I. Zborovsk\'y}, 
  Phys.\ Lett.\  {\bf B206}  (1988) 159;
{M. Pachr, V. \u{S}im\'ak, M. \u{S}umbera and I. Zborovsk\'y}, 
  Mod.\ Phys.\ Lett.\  {\bf A77}   (1992) 2333;
{M. \u{S}umbera}, talk at the Alice PPR meeting, CERN, April 2001.

\bibitem{TH:hqlett:2}
{A.~Giovannini, S.~Lupia and R.~Ugoccioni}, Phys.\ Lett.\  {\bf B374}
  (1996)  231.

\bibitem{TH:L3:mangeol}
{P. Achard et al, (L3 Collaboration)}, preprint CERN-EP/2001-072
  (hep-ex/0110072), CERN.

\bibitem{TH:Bjorken2}
{J.D.~Bjorken}, Phys.\ Rev.\  {\bf D47}   (1993) 101.

\bibitem{TH:pzero}
{S. Lupia, A. Giovannini and R. Ugoccioni}, Z. Phys.\ {\bf C66}   (1995) 195.

\bibitem{TH:VanHove:ptQGP}
{L. Van Hove}, Phys.\ Lett.\  {\bf B118}   (1982) 138.


\end{thebibliography}
\end{document}